\documentclass[article,twocolumn,amsmath,floats,showpacs,nofootinbib]{revtex4}

\usepackage{graphicx}

\usepackage{textcomp}
\usepackage{hyperref}

\hypersetup{colorlinks=true}

\begin{document}

\title{Magnetoquantum oscillations in the resistance of metallic point
contacts: Influence of nonequilibrium phonons}

\author{N. L. Bobrov$^{1,2,}$\footnote{Corresponding author.},  J. A. Kokkedee$^1$,  N. N. Gribov$^{1,2}$, I. K. Yanson$^2$, A. G. M. Jansen$^1$,  P. Wyder$^1$}

\affiliation
{$^1$High Magnetic Field Laboratory, Max-Planck Institut fur Festkorperforschung and Centre National de la Recherche Scientifique, F-38042, Grenoble, Cedex 09, France\\$^2$B.I.~Verkin Institute for Low Temperature Physics and Engineering, of the National Academy of Sciences of Ukraine, prospekt Lenina, 47, Kharkov 61103, Ukraine\\
E-mail address: bobrov@ilt.kharkov.ua}
\published {\href{http://www.sciencedirect.com/science/article/pii/0921452695005544}{Physica B} \textbf{218}, 42(1996)}
\date{\today}

\begin{abstract}The amplitude of magnetoresistance quantum oscillations of $Al$ and $Be$ point contacts in a magnetic field parallel to
the contact axis has been studied as a function of voltage applied over the contact. It was found that for one group of
contacts the oscillation amplitude nonmonotonously increases with the bias voltage increase, while for the other part of
the contacts a decrease of amplitude was observed. The scattering of electrons with nonequilibrium phonons and also
phonon-phonon collisions will be discussed as the possible reasons of the observed effects.
\end{abstract}
\maketitle

\section{INTRODUCTION}
The resistance of metallic point contacts at low temperatures
shows the oscillations as a function of magnetic
field, which are due to the Landau quantization of the
conduction electron energy spectrum \cite{Gribov, Swartjes}. An increase
in temperature reduces the amplitude of oscillations. The
corresponding reduction factor is
\begin{equation} \label{eq__1}
{{R}_{T}}=\frac{2{{\pi }^{2}}n{{k}_{B}}T/\hbar {{\omega }_{c}}}{\operatorname{sh}(2{{\pi }^{2}}n{{k}_{B}}T/\hbar {{\omega }_{c}})};
\end{equation}
where $n$ is the harmonic number in the oscillations and
${{\omega }_{c}}=eB/m^{*}c$ the cyclotron frequency. Elastic scattering
of the electrons by impurities also causes the decrease of
the oscillation amplitude nearly the same as resulted
from the temperature increase from actual value $T$ up to
${{T}_{eff}}=T+x$. The reduction factor (the Dingle factor) has the form

\begin{equation} \label{eq__2}
{R}_{D}=\exp (-2{{\pi }^{2}}{{k}_{B}}nx/\hbar {{\omega }_{c}}),
\end{equation}
where $x=\hbar /2{{\pi }}{{k}_{B}}\tau $ is the Dingle temperature, and  $\tau$ is the elastic lifetime of the electrons. As follows from Eqs. \eqref{eq__1}
and \eqref{eq__2} the ratio between the first and the second harmonic
of the oscillations is
\begin{equation} \label{eq__3}
{{A}_{1}}/{{A}_{2}}=C\text{ch}(2{{\pi }^{2}}{{k}_{B}}/\hbar {{\omega }_{c}})\exp (\pi /{{\omega }_{c}}\tau ),
\end{equation}
where the constant $C$ has a meaning of the ratio of the
harmonics at $T=0$ in the absence of scatterers. From
the theory for noninteracting particles it follows that the
Dingle temperature can be affected by any electron scattering
process, including electron-phonon collisions \cite{Shoenberg}.
At the same time, with many-body interactions taken
into account, the scattering of electrons on phonons
seems to have no effect on the Dingle temperature \cite{Shoenberg}.
The study of the influence of electron-phonon collisions
on the amplitude of magnetoquantum oscillations in
traditional experiments is limited to metals with low
Debye energy, in which phonons can be excited at a tem,
perature low enough to observe the effect. Metallic point
contacts present a unique opportunity to study the influence
of nonequilibrium phonons generated in the contact
on the quantum oscillations. Just by applying the bias
over the contact, the nonequilibrium phonons can be
generated in the constriction with any possible energy up
to $eV$ \cite{Kulik}. In this work we study the effect of electron-
phonon scattering on the amplitude of the resistance
oscillations of $Al$ and $Be$ point contacts in a
magnetic field parallel to the contact axis.
\section{Experimental details}
Point contacts were formed between edges of two
single crystal electrodes with the same crystallographic
orientations. The applied magnetic field was always parallel
to the contact axis. All the experiments were carried
out at the temperature $1.3\ K$.
\begin{figure}[]
\includegraphics[width=8cm,angle=0]{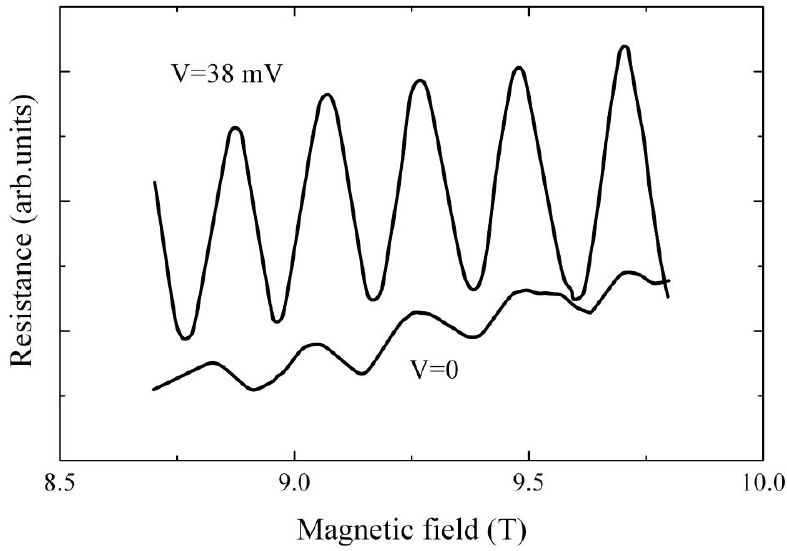}
\caption[]{Fragments of magnetoresistance oscillations of an $Al$
point contact of $R_0=1.41\ \Omega$ for two different bias voltages.}
\label{Fig1}
\end{figure}
The first and the second
derivatives of the $I-V$ curve were measured by standard
modulation technique. To determine the energy dependence
of the amplitude of magnetoresistance oscillations
the following procedure was used. For several fixed bias
voltage $dV/dI(B)$ dependencies were recorded in the
same interval of magnetic field (usually in between 8 and
10 $T$). Fig. \ref{Fig1} shows the fragments of $dV/dI(B)$ curves for
an $Al$ point contact for two bias voltages. The amplitude
of oscillations $A_1$ and its second harmonic $A_2$ was obtained
by means of Fourier analysis. For one group of
contacts the voltage dependence $A_1(eV)$ was found to
vary nonmonotonously with applied voltage with similarities
to the point contact spectrum of electron-phonon
interaction. Such an effect was observed for low Ohmic
$Al$ point contacts (Fig. \ref{Fig2}(a)) and $Be$ contacts (Fig. \ref{Fig3}(a)).
\begin{figure}[]
\includegraphics[width=8.7cm,angle=0]{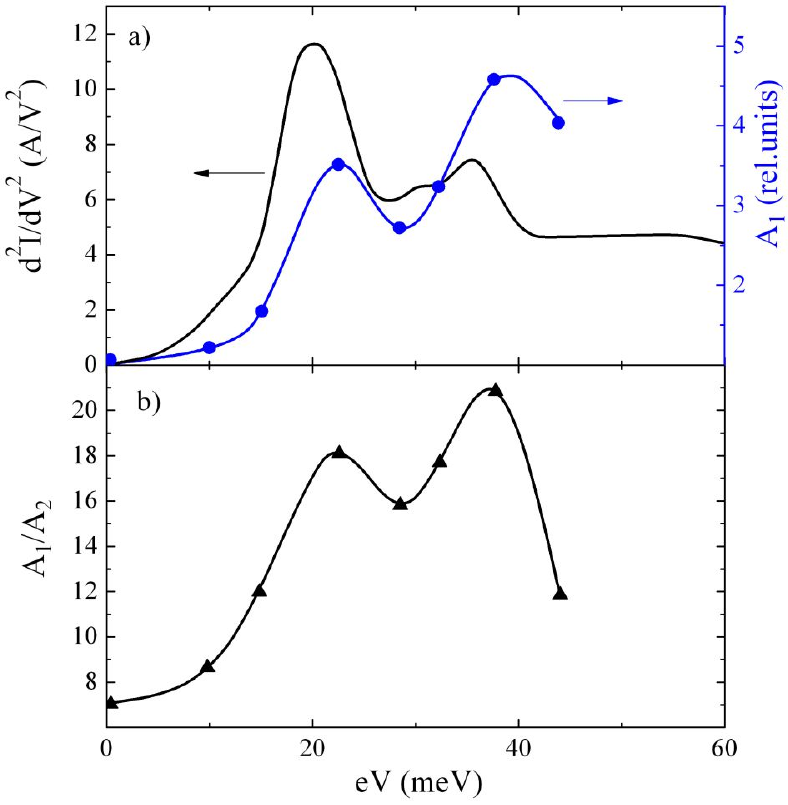}
\caption[]{(a) Point contact spectrum of an $Al$ point contact of
 ${{R}_{0}}=1.41\ \Omega$ and relative magnitude of the first harmonic of the
resistance oscillations versus bias voltage; $\Delta R/{{R}_{0}}\approx 5\times {{10}^{-4}}$ at
$V = 0$ and $B = 9.5\ T$.
 \\(b) Ratio between the amplitude of the first
and the second harmonic of the magnetoresistance oscillations
versus bias voltage.}
\label{Fig2}
\end{figure}
\begin{figure}[]
\includegraphics[width=8.7cm,angle=0]{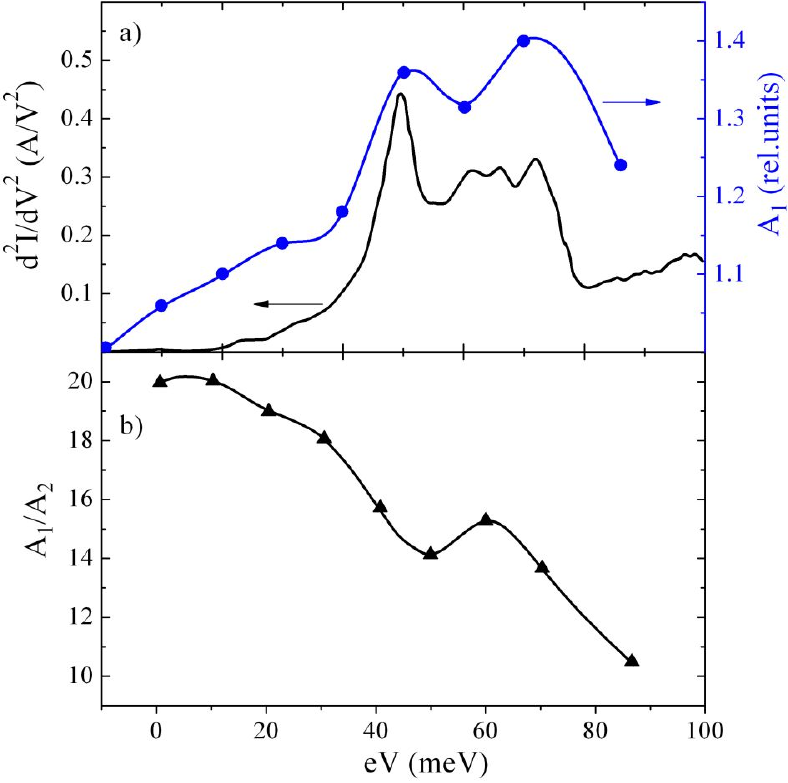}
\caption[]{(a) Point contact spectrum of an $Be$ point contact of
 ${{R}_{0}}=8.66\ \Omega$ and relative magnitude of the first harmonic of the
resistance oscillations versus bias voltage; $\Delta R/{{R}_{0}}\approx 2\times {{10}^{-3}}$ at
$V = 0$ and $B = 9.5\ T$.
 \\(b) Ratio between the amplitude of the first
and the second harmonic of the magnetoresistance oscillations
versus bias voltage.}
\label{Fig3}
\end{figure}
High Ohmic $Al$ point contacts show a decrease of the
amplitude of oscillations in their magnetoresistance with
the bias voltage increase (Fig. \ref{Fig4}(a)).
\begin{figure}[]
\includegraphics[width=8.7cm,angle=0]{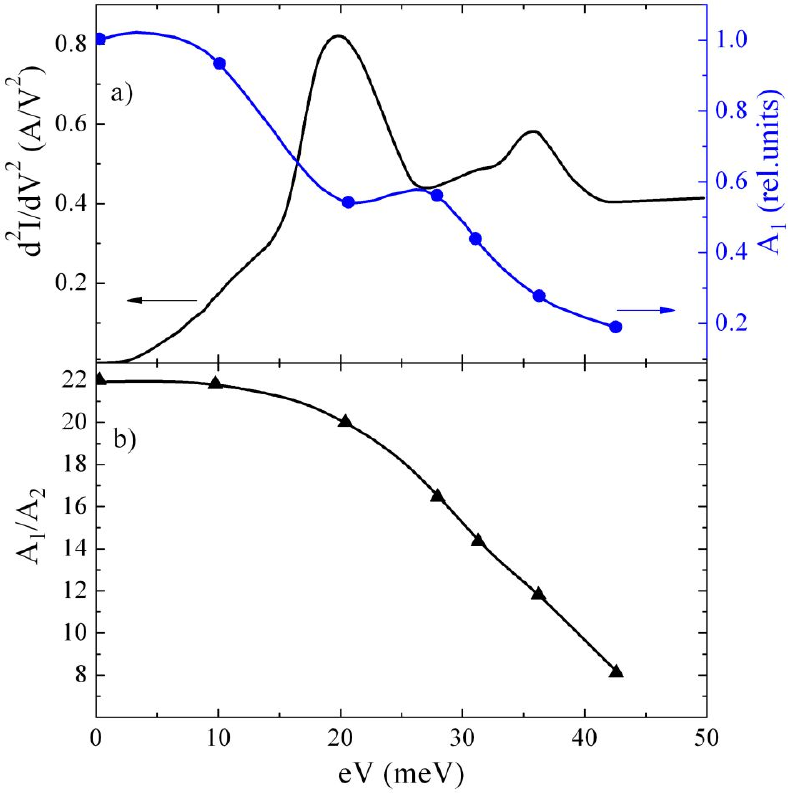}
\caption[]{(a) Point contact spectrum of an $Al$ point contact of
 ${{R}_{0}}=8.73\ \Omega$ and relative magnitude of the first harmonic of the
resistance oscillations versus bias voltage; $\Delta R/{{R}_{0}}\approx 3\times {{10}^{-3}}$ at
$V = 0$ and $B = 9.5\ T$.
 \\(b) Ratio between the amplitude of the first
and the second harmonic of the magnetoresistance oscillations
versus bias voltage.}
\label{Fig4}
\end{figure}
\section{Discussion}
Phonons generated in point contact are the result of
the energy relaxation of electrons injected from the contact.
In analogy with elastic impurity scattering, the scattering
of electrons on these nonequilibrium phonons will
influence the amplitude of magnetoquantum oscillations.
As with the elastic scattering \cite{Bogachek}, phonon-induced scattering
within the constriction region will cause the
randomization of the momenta of electrons. This randomization
will lead to an effective increase of the $z$-component
of the transport velocity of electrons from the
extremal cross section of the Fermi surface and correspondingly
to an increase of quantum oscillations in the
point contact magnetoresistance. At the same time, the
scattering of electrons on nonequilibrium phonons in the
boundaries of the point contact will cause the increase of
the Dingle temperature and thus a decrease of the oscillation
amplitude. The resulting effect will be determined by
the balance between these two processes.

It is obvious that the purity of the metal in the region
adjacent to the constriction area is an important factor
affecting the oscillations of the electron density of states.
As follows from Eq. \eqref{eq__2} the oscillation amplitude has
nonlinear dependence on the concentration of the impurity
scatterers. In the case of a large value of $\omega_{c}\tau$ (low
impurity concentration) it saturates and exponentially
drops at small values of $\omega_{c}\tau$. Hence, if the concentration
of defects in the point contact boundaries is low, the small increase in the number of the electron-phonon
scattering events leads to insignificant decrease of the
oscillation amplitude. However, if the impurity concentration
in the boundaries is high enough, a similar increase
in the number of electron-phonon collisions may
dramatically affect the oscillations.

Taking into account all the factors mentioned above
one can expect an increase of the amplitude of magnetoresistance
oscillations in relatively large (compare to
the diffusive mean free path for phonons) low Ohmic
contacts with pure boundaries. Nonequilibrium phonons
generated in such contacts are accumulated within the
constriction area and play an effective part in the isotropization
of the electron momenta. Purity of the boundaries
is of great importance because the oscillations are mainly
the result of quantizing conditions there. Fig. \ref{Fig2}(a) shows
the example of $A_{1}(eV)$ dependence for one of low Ohmic
$Al$ point contact. The oscillation amplitude dramatically
increases with the bias voltage increase. A low value of
the ratio $A_{1}/A_{2}$ at $eV = 0$ (Fig. \ref{Fig2}(b)) indicates the high
purity of the boundaries of this contact. In the case of
small (high Ohmic) contacts the generated phonons leave
the constriction area. If, additionally, the boundaries of
the contact are not pure enough (high value of $A_{1}/A_{2}$),
the decrease of the amplitude of magnetoresistance oscillations
is observed (see Fig. \ref{Fig4}).

As for $Be$, the increase of the oscillation amplitude with
bias rise was observed for all contacts (Fig. \ref{Fig3}). This is
probably due to high rigidity of the $Be$ crystal lattice, which
prevents the spreading of the deformation to the boundaries
during the contact formation. These deformations are concentrated
near the surface, forming the effective reflectors,
which return the phonons back to the contact.

The given model can not only explain the monotonous
part of the voltage dependence but also its similarity to
the electron phonon interaction spectrum. The group
velocity of the phonons generated by the electron flow
depends on their energy and is minimum for phonons
with energies close to the maxima of the phonon density
of states. This means that the phonon energy relaxation
length will decrease stepwise at voltages corresponding
to the characteristic phonon frequencies. Therefore, one
should expect the similarity of $A_{1}(eV)$ with the first
derivative $dV/dI(eV)$ of the $I-V$ curve, which is not the
case in our experiments. However, in these speculations
the role of phonon-phonon collisions was not taken into
account.

The stepwise decrease of the phonon energy relaxation
length and correspondingly their group velocity at voltages
higher than or of the order of typical phonon energies
leads to a significant increase of the frequency of
phonon-phonon collisions. This will result in the thermalization
of phonons, which may have the following
consequences:
\begin{enumerate}
  \item {Phonon-phonon collisions will decrease the number
of high-frequency phonons, which are the most efficient
electron scatterers because of their small group
velocities, while the thermalized phonons will leave the
contact area.}
  \item {An increase of the total number of phonons in the
contact boundaries, resulting from thermalization leads
to a stronger suppression of oscillations in this area.}
\end{enumerate}

Hence, one can expect that for bias voltages higher
than or of the order of characteristic phonon energies, the
rise of $A_{1}(eV)$ will slow down or even a decrease may be
observed.

Now consider the bias dependence of the ratio between
the first and the second harmonics of the oscillations
$A_{1}/A_{2}(eV)$. From Eq. \eqref{eq__3} it can be readily seen that the
magnitude of $A_{1}/A_{2}$ depends on the value of the electron
mean free path in the region of metal where the quantization
of electron density of states takes place. Due to
geometry of experiments (magnetic field direction is parallel
to the contact axis) the ratio $A_{1}/A_{2}$ at $eV = 0$ contains
the information about electron mean free path in
the region adjacent to the constriction area. As for the
energy dependence of $A_{1}/A_{2}$ we assume the existence of
two mechanisms, which can be responsible for the behaviour
observed:

\begin{enumerate}
  \item {A decrease of (quasi-)elastic mean free path of electrons
in the contact boundaries due to the scattering of
electrons on nonequilibrium phonons.}
  \item {Redistribution of the relative contribution to the
information about electron mean free path from the areas
with different concentrations of impurities. If the impurity concentration is homogeneous over the entire contact
region, the ratio $A_{1}/A_{2}$ will increase with the bias
rise. If, however, the concentration of impurities decreases
with the distance to the contact centre, the opposite
effect will be observed.}
\end{enumerate}

Certainly, coexistence of both mechanisms is possible.
It is obvious that the first of them will lead to an increase
of the $A_{1}/A_{2}$ ratio only (see Eq. \eqref{eq__2}).

In summary, we have studied the energy dependence of
the oscillation amplitude in magnetoresistance of metallic
point contacts in a magnetic field parallel to the
contact axis. It was found for the first time that this
dependence shows a nonmonotonous behaviour and in
some cases its shape is similar to the point contact spectrum
of electron phonon interaction. The additional
scattering of quantized electrons on the phonons generated
by the flow of electrons through the contact may be
the possible reason of the observed effect.

\end{document}